\documentclass[aps,prl,amsmath,amssymb,reprint,twocolumn,superscriptaddress,showpacs]{revtex4-1}
\usepackage{epsfig,amssymb}
\usepackage{graphicx}
\usepackage{dcolumn}
\usepackage{bm}
\usepackage{dcolumn}
\usepackage{multirow}
\begin{document}
\title{First-principles theory of electron-spin fluctuation coupling
  and superconducting instabilities in iron selenide}

\author{Johannes~Lischner}
\email{jlischner@civet.berkeley.edu}
\affiliation{Department of Physics, University of California,
  Berkeley, California 94720, USA, and Materials Sciences Division,
  Lawrence Berkeley National Laboratory, Berkeley 94720, USA.}
\author{Timur~Bazhirov}
\affiliation{Department of Physics, University of California,
  Berkeley, California 94720, USA, and Materials Sciences Division,
  Lawrence Berkeley National Laboratory, Berkeley 94720, USA.}
\author{Allan H. MacDonald}
\affiliation{Department of Physics, The University of Texas at Austin,
  Austin, Texas, 78712, USA.}
\author{Marvin L. Cohen}
\affiliation{Department of Physics, University of California,
  Berkeley, California 94720, USA, and Materials Sciences Division,
  Lawrence Berkeley National Laboratory, Berkeley 94720, USA.}
\author{Steven G. Louie}
\affiliation{Department of Physics, University of California,
  Berkeley, California 94720, USA, and Materials Sciences Division,
  Lawrence Berkeley National Laboratory, Berkeley 94720, USA.}

\begin{abstract}
  We present first-principles calculations of the coupling of
  quasiparticles to spin fluctuations in iron selenide and discuss
  which types of superconducting instabilities this coupling gives
  rise to. We find that strong antiferromagnetic stripe-phase spin
  fluctuations lead to large coupling constants for superconducting
  gaps with $s_\pm$-symmetry, but these coupling constants are
  significantly reduced by other spin fluctuations with small wave
  vectors. An accurate description of this competition and an
  inclusion of band structure and Stoner parameter renormalization
  effects lead to a value of the coupling constant for an
  $s_\pm$-symmetric gap which can produce a superconducting transition
  temperature consistent with experimental measurements.
\end{abstract}

\maketitle

\emph{Introduction}.---The discovery of superconductivity in
iron-based compounds with transition temperatures higher than 50
Kelvin in 2008 \cite{kamihara2008iron} has generated considerable interest
in recent years and led to intense research activity. These
materials consist of iron atoms with additional pnictogen or chalcogen
atoms located above and below the plane of the iron atoms. Additional
intercalating layers give rise to various families of iron-based
compounds, such as the 1111-, 122-, 111-families
\cite{stewart2011superconductivity}. Iron selenide (FeSe) belongs to
the structurally simplest family, the 11-family, which does not
contain a spacer layer between the iron-selenium layer and exhibits
superconductivity with transitions temperatures up to 37 Kelvin under
pressure \cite{medvedev2009electronic,imai2009does}. Recently, there
has also been considerable interest in the properties of FeSe
monolayers, where superconductivity with transition temperatures
exceeding 60 Kelvin has been reported
\cite{qing2012interface,ge2014superconductivity}.

An important question in these materials is the nature of the
microscopic pairing mechanism and the
symmetry of the superconducting gap. As in the cuprates,
superconductivity typically emerges in the iron-based compounds when
an antiferromagnetic parent state is doped. This observation led 
to the proposal \cite{Mazin} that spin fluctuations (paramagnons) could
act as the superconducting glue in these materials
\cite{Mazin}. However, in contrast to the cuprates, the iron-based
materials generally have multiple Fermi surfaces with several
electron pockets at the M-point of the Brillouin zone (corresponding
to the unit cell containing two iron atoms) and multiple hole pockets
at the $\Gamma$-point. Mazin \emph{et al.} suggested that
spin-fluctuation mediated scattering of Cooper pairs between electron
pockets and hole pockets gives rise to an $s_{\pm}$-symmetric
superconducting gap \cite{Mazin}, which has a constant absolute
magnitude, but switches sign between electron and hole pockets.  The
sign-changing gap was predicted to give rise to a resonance in the
neutron scattering spectrum \cite{maier2009neutron}, which was
subsequently observed in several experiments
\cite{inosov2009normal,babkevich2011spin}.

Many theoretical approaches have been developed to study
spin-fluctuation mediated superconductivity in the iron-based
compounds. In purely empirical approaches, both the electronic band
structure and the interacting spin susceptibility are parametrized
using experimental data (such as angle-resolved photoemission, nuclear
magnetic resonance and neutron scattering
results)\cite{Eschrig,fernandes2013suppression}. In another approach
\cite{Aoki,graser2009near}, a theoretical band structure from a
density-functional theory calculation is used to parametrize a
tight-binding Hamiltonian with added interaction parameters (such as the
Hubbard $U$ or Hund's $J$) adjusted to reproduce experimental
findings. Then, superconducting properties of the resulting
Hamiltonian are studied.

While the aforementioned theories have been very instructive, their
applications have been limited by the availability of concrete
experimental data needed to determine their input parameters. Hence,
there is a need for a fully first-principles theory without empirical
parameters. We have recently developed such an \emph{ab initio} theory
\cite{lischner2014effect} for the spin fluctuation-electron coupling
based on the work of Overhauser and coworkers on the homogeneous
electron gas \cite{OverhauserKukkonen,OverhauserZhu}.

In this paper, we apply our first-principles theory of the
electron-spin fluctuation coupling to iron selenide, the structurally
simplest iron-based superconductor. We reveal a complex interplay
between different spin fluctuations. In agreement with experiment,
superconductivity depends sensitively on the height of the selenium
atoms relative to the iron atom plane. In the vicinity of a transition
to an antiferromagnetic stripe phase, superconductivity with an
$s_{\pm}$-symmetric gap function becomes favorable.

\emph{Methods}.---The superconducting order parameter, the gap
function, is typically expressed as $\Delta(\bm{k})=|\Delta|g(\bm{k})$
with $g(\bm{k})$ describing the symmetry of the gap function
\cite{maier2009origin}. To investigate which types of superconducting
instabilities an effective spin-fluctuation-mediated
electron-electron interaction $V_{SF}$ gives rise to, we compute the
electron-spin fluctuation coupling strength for different symmetry
functions $g(\bm{k})$ according to \cite{maier2009origin}
\begin{align}
 & \lambda[g] = -D_F  \times \\
 & \frac{\sum_{n\bm{k},n'\bm{k}'}
    g(\bm{k})V_{SF}(n\bm{k},n'\bm{k}')g(\bm{k}')\delta(\epsilon_{n\bm{k}}-\epsilon_F)\delta(\epsilon_{n'\bm{k}'}-\epsilon_F)}{[\sum_{n\bm{k}}
      g^2(\bm{k}) \delta(\epsilon_{n\bm{k}}-\epsilon_F)]^2},
    \label{eq:lambda}
\end{align}
where $\epsilon_F$ denotes the Fermi energy and $D_F$ is the density
of states per spin at the Fermi energy. We define
$V_{SF}(n\bm{k},n'\bm{k}')$ as the matrix element of the
spin-fluctuation mediated interaction for the scattering of the
spin-singlet Cooper pair $(n\bm{k}\uparrow,n-\bm{k}\downarrow)$ to
$(n'\bm{k}'\uparrow,n'-\bm{k}'\downarrow)$ (with $n$ denoting a band
and $\bm{k}$ a $\bm{k}$-point in the Brillouin zone). We follow the
standard convention \cite{maier2009origin} and normalize $g(\bm{k})$,
such that $D_F = \sum_{n\bm{k}} g^2(\bm{k})
\delta(\epsilon_{n\bm{k}}-\epsilon_F)$.

The effective spin-fluctuation mediated interaction is obtained using
a recently developed first-principles formalism
\cite{lischner2014effect,schweflinghaus2014renormalization} based on
the work of Overhauser \emph{et al.} for the homogeneous electron
gas \cite{OverhauserKukkonen,OverhauserZhu}. In this approach, the
effective interaction is expressed as the sum of a bare Coulomb
interaction, a contribution arising from charge fluctuations and a
contribution arising from spin fluctuations, which is given by
\begin{align}
 & V_{SF}(\bm{r},\bm{r}',\omega)=3 \times \nonumber \\
&\int d\bm{r}_1 \int d\bm{r}_2 I_{xc}(\bm{r},\bm{r}_1) \chi_S(\bm{r}_1,\bm{r}_2,\omega) I_{xc}(\bm{r}_2,\bm{r}'),
\end{align}
with $I_{xc}(\bm{r},\bm{r}')=\delta^2 E_{xc}/[\delta m(\bm{r}) \delta
m(\bm{r}')]$ ($E_{xc}$ is the exchange-correlation energy and
$m(\bm{r})$ the spin density). Also, $\chi_S(\bm{r},\bm{r}',\omega)$ denotes the
interacting spin susceptibility of a nonmagnetic system, which is
obtained by solving the Dyson-like equation of time-dependent
density-functional theory \cite{essenberger2012paramagnons}
\begin{align}
 & \chi_S(\bm{r},\bm{r}',\omega) = \chi_0(\bm{r},\bm{r}',\omega) \nonumber \\
&+\int d\bm{r}_1 \int d\bm{r}_2\chi_0(\bm{r},\bm{r}_1,\omega) I_{xc}(\bm{r}_1,\bm{r}_2) \chi_S(\bm{r}_2,\bm{r}',\omega),
\end{align}
with $\chi_0$ denoting the non-interacting susceptibility. 

The physically appealing effective spin fluctuation-mediated
interaction of Overhauser \emph{et al.} can be put on a firm
theoretical footing by an analysis of Feynman diagrams
\cite{VignaleSingwi,essenberger2014superconducting}. While this theory
neglects certain diagrams corresponding to nonlinear polarization
processes, it should be valid in the vicinity of the Fermi surface and
thus describe superconducting properties accurately.

\emph{Computational details}.---To obtain a mean-field theory starting
point for the calculation of the spin-fluctuation mediated
interaction, we carry out density-functional theory calculations in a
plane-wave basis using the Quantum ESPRESSO program package
\cite{QuantumEspresso}. We employ the local-density approximation
(Perdew-Zunger parametrization) and norm-conserving pseudopotentials
with a 55 Ry energy cutoff. For iron, the non-linear core correction
is used. We use experimental lattice constants ($a=3.77$~\AA,
$c=5.52$~\AA) \cite{phelan2009neutron} of the tetragonal phase. We
have carefully verified that our band structures and magnetic phase
diagrams agree with all-electron results
\cite{aichhorn2010theoretical,kumar2012chalcogen}.

We then compute the non-interacting susceptibilities on a $32 \times 32
\times 1$ $\bm{k}$-point grid in the Brillouin zone using 50 empty states and
a plane-wave cutoff of 45 Ry. For this, we use the BerkeleyGW program
package \cite{BGWpaper}. Next, we compute the interacting spin
susceptibilities. Within the local-density approximation, we need to
evaluate
$I^{LSDA}_{xc}(\bm{r},\bm{r}')=f(n(\bm{r}))\delta(\bm{r}-\bm{r}')$
with $n(\bm{r})$ being the ground-state density (including the core
contribution) and $f(n)$ is obtained analytically by computing the
second derivative of the exchange-correlation energy with respect to
the spin density. Finally, the Fermi surface averages in
Eq.~\eqref{eq:lambda} are evaluated using a Lorentzian representation
of the $\delta$-function [$\delta_\eta(\omega) = 1/\pi \times
\eta/(\omega^2 + \eta^2)$] with $\eta=0.1$~eV.

\begin{figure}
  \includegraphics[width=8.cm]{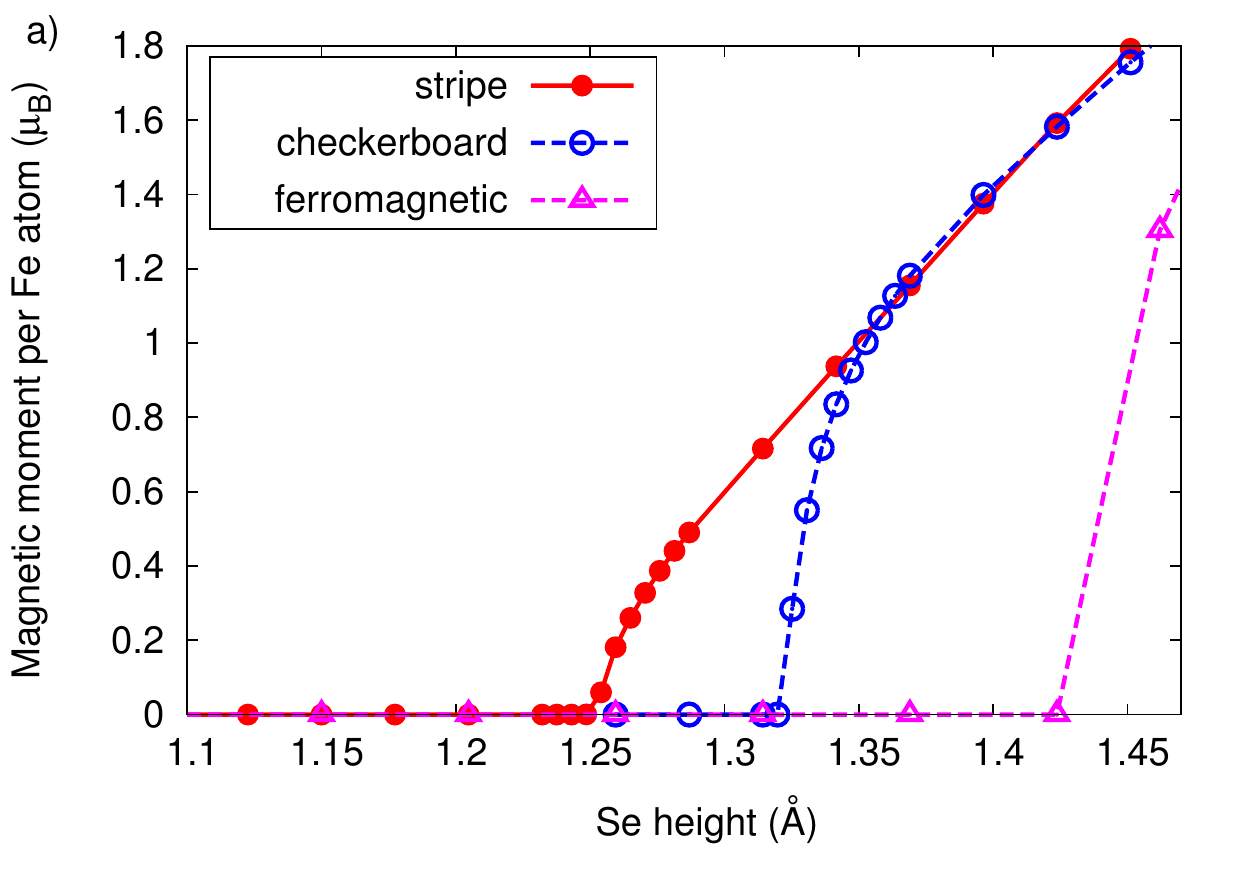}
  \includegraphics[width=8.cm]{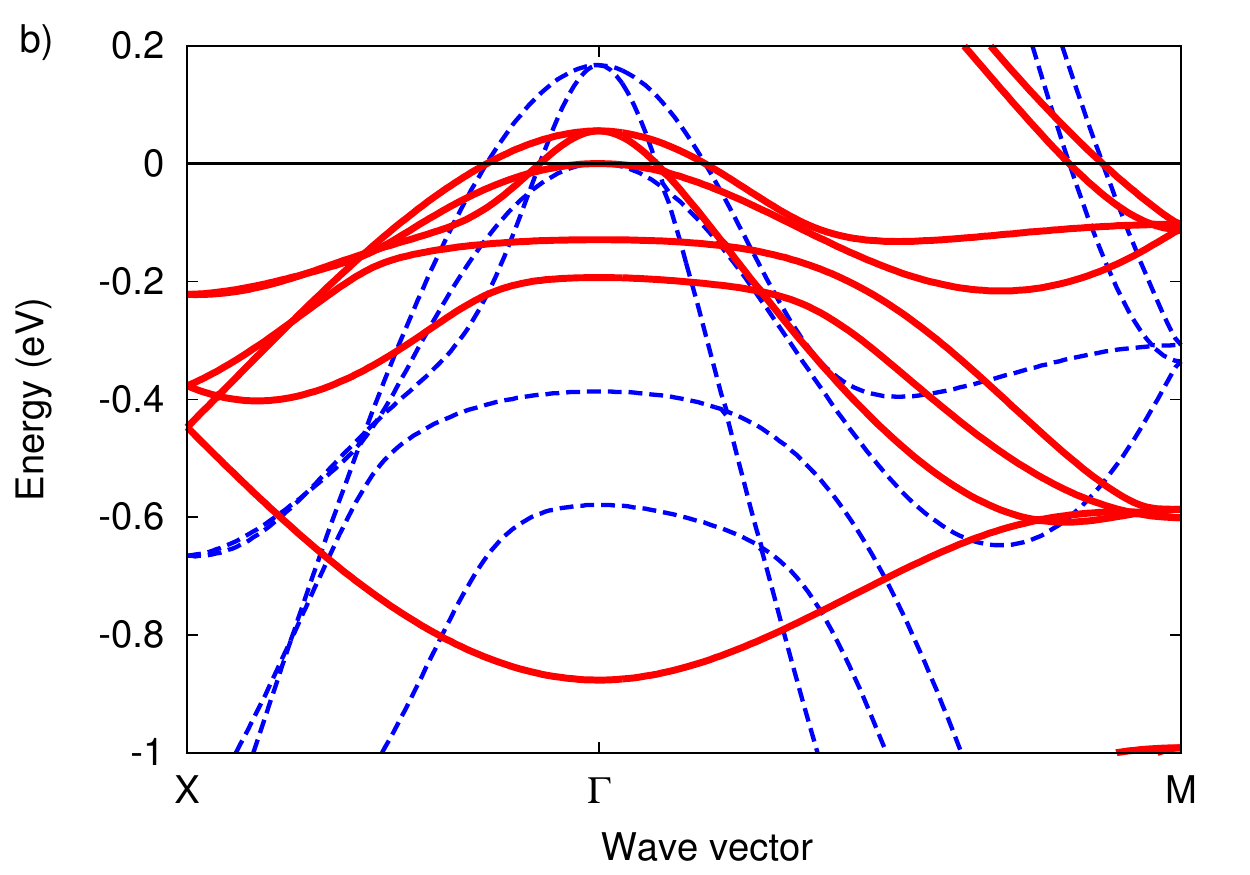}  
  \caption{(a): Magnetic moment per iron atom in the stripe
    antiferromagnetic phase (red filled dots), checkerboard
    antiferromagnetic phase (blue empty dots) and ferromagnetic phase
    (magenta diamonds) as function of the height of the selenium
    atoms above (and below) the plane of the iron atoms. (b) DFT-LDA
    band structure of iron selenide at the experimental selenium
    height (dashed blue lines) and the renormalized band structure
    (solid red lines), where all energies have been divided by a factor of
    3 \cite{tan2013interface}.}
  \label{fig:bands}
\end{figure}

\emph{Results---}. Figure~\ref{fig:bands}(a) shows the magnetic
moments of various magnetic phases of FeSe as a function of the
selenium height, which has been identified as a crucial parameter for
the occurence of superconductivity in this system and other iron-based
compounds \cite{okabe2010pressure}. At low selenium heights, the
system is nonmagnetic. As the height is increased, a transition to an
antiferromagnetic stripe phase occurs. At even higher Se heights, an
antiferromagnetic checkerboard phase also has a lower energy than the
nonmagnetic state and a moment develops. Finally, at Se heights larger
than 1.4~\AA~a ferromagnetic moment is observed. Note that bulk
FeSe is experimentally known to be nonmagnetic
\cite{medvedev2009electronic}, while a recent ARPES experiment
reported indications that multilayer FeSe is in a stripe phase state
\cite{tan2013interface}.

Figure~\ref{fig:bands}(b) shows the DFT-LDA band structure of FeSe at
the experimental selenium height of 1.47~\AA \cite{phelan2009neutron}.
We observe three hole pockets in the vicinity of the $\Gamma$-point
and two electron pockets near the M-point (note that a Brillouin zone
corresponding to a unit cell containing \emph{two} iron atoms is
used). While DFT-LDA band structures agree qualitatively with
experimental ARPES measurements for many iron-based compounds, the
effective masses of the electron and hole pockets near the Fermi level
are typically underestimated by a factor of 2 or 3
\cite{lu2008electronic,evtushinsky2014strong}. A similar finding was
reported in a recent ARPES experiment on FeSe multilayers
\cite{tan2013interface}. In Fig.~\ref{fig:bands}(b) we also show the
renormalized band structure obtained by dividing all DFT-LDA energies
by a factor of 3. We note that a recent Shubnikov-de Haas oscillation
measurement reported significantly reduced Fermi surfaces
\cite{terashima2014anomalous}. Further experimental work is necessary
to fully resolve this issue.

\begin{figure}
  \begin{tabular}{c c}
  \includegraphics[width=4.cm]{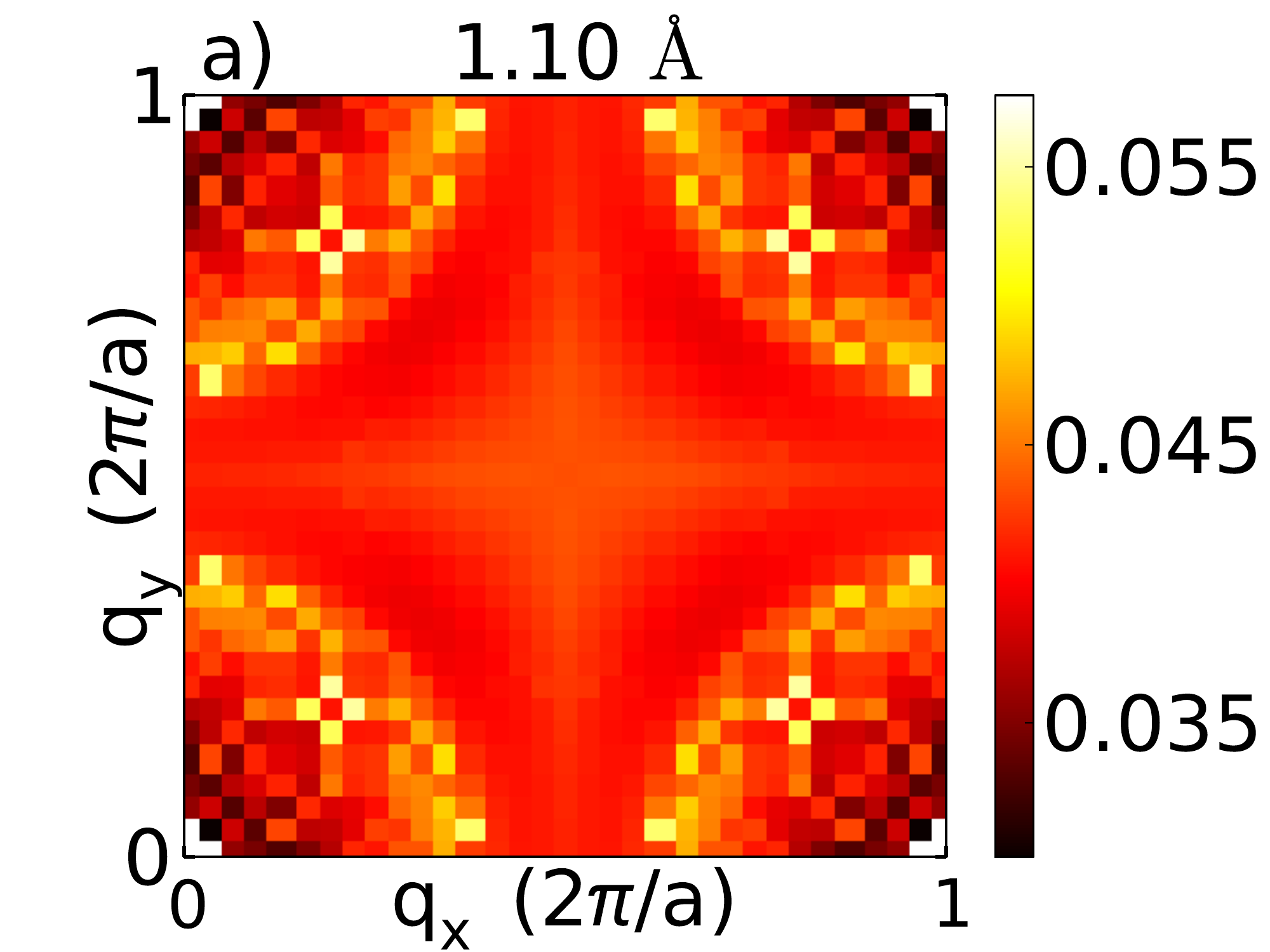} &
  \includegraphics[width=4.cm]{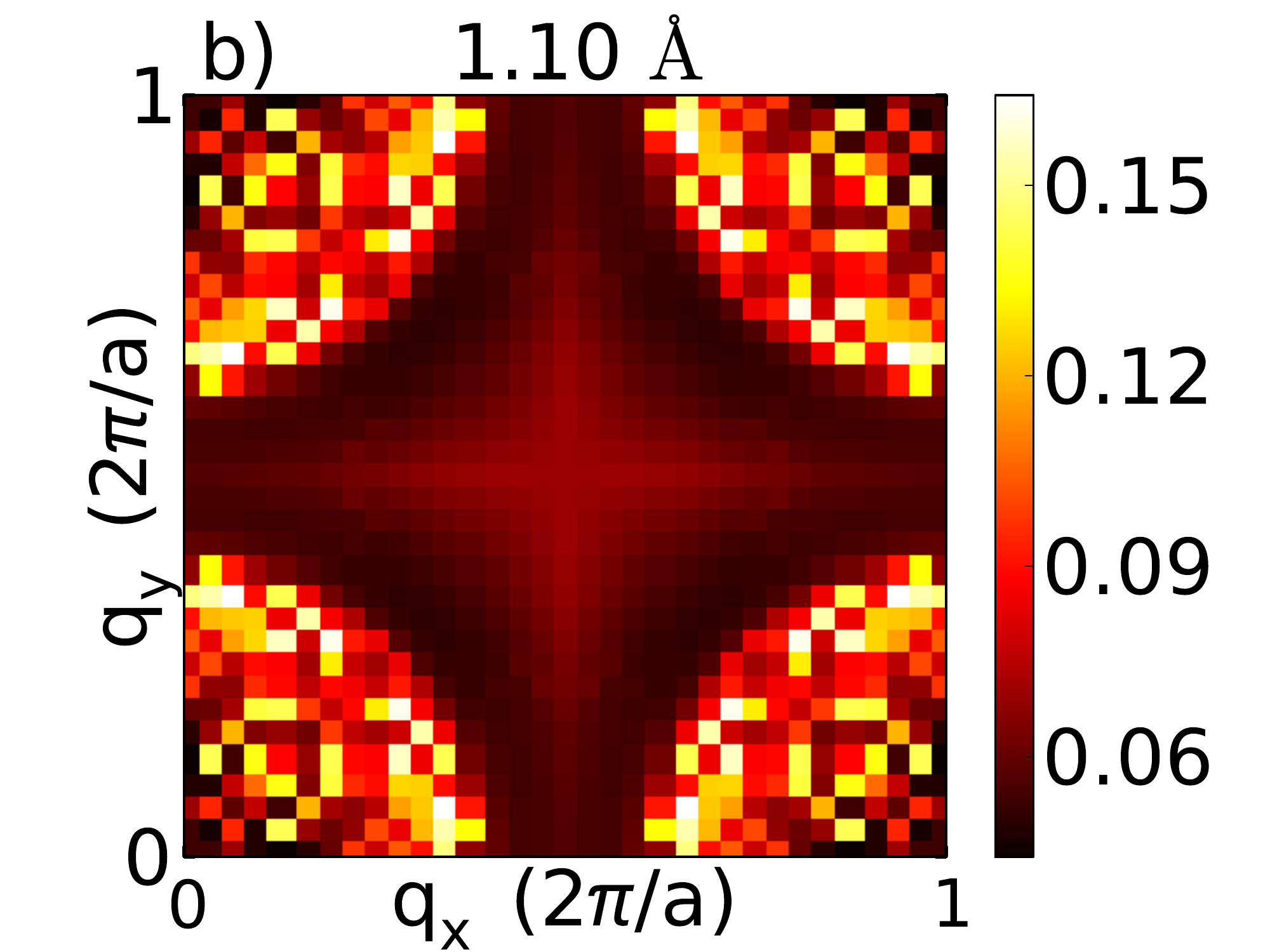} \\
    \includegraphics[width=4.cm]{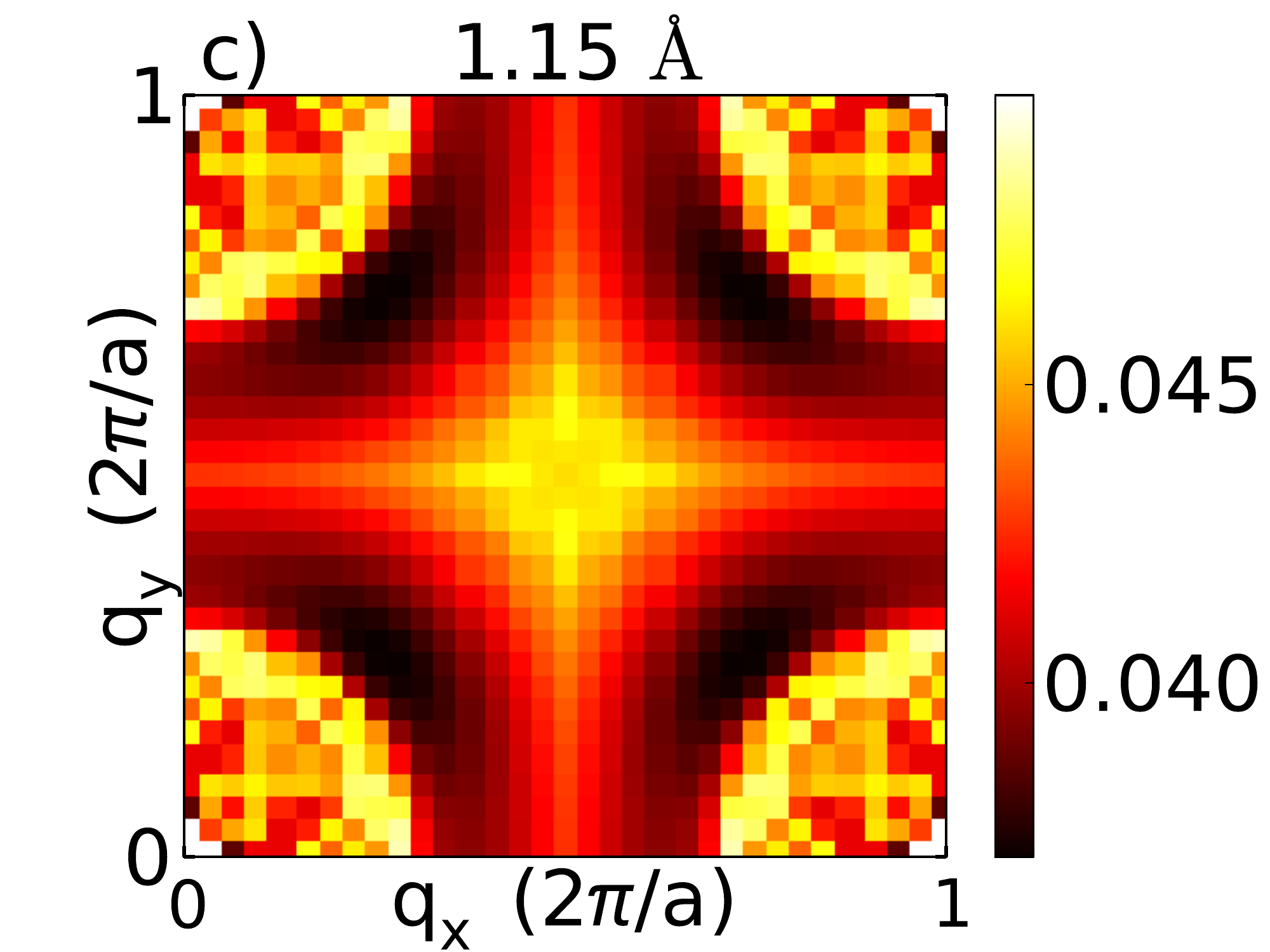} &
  \includegraphics[width=4.cm]{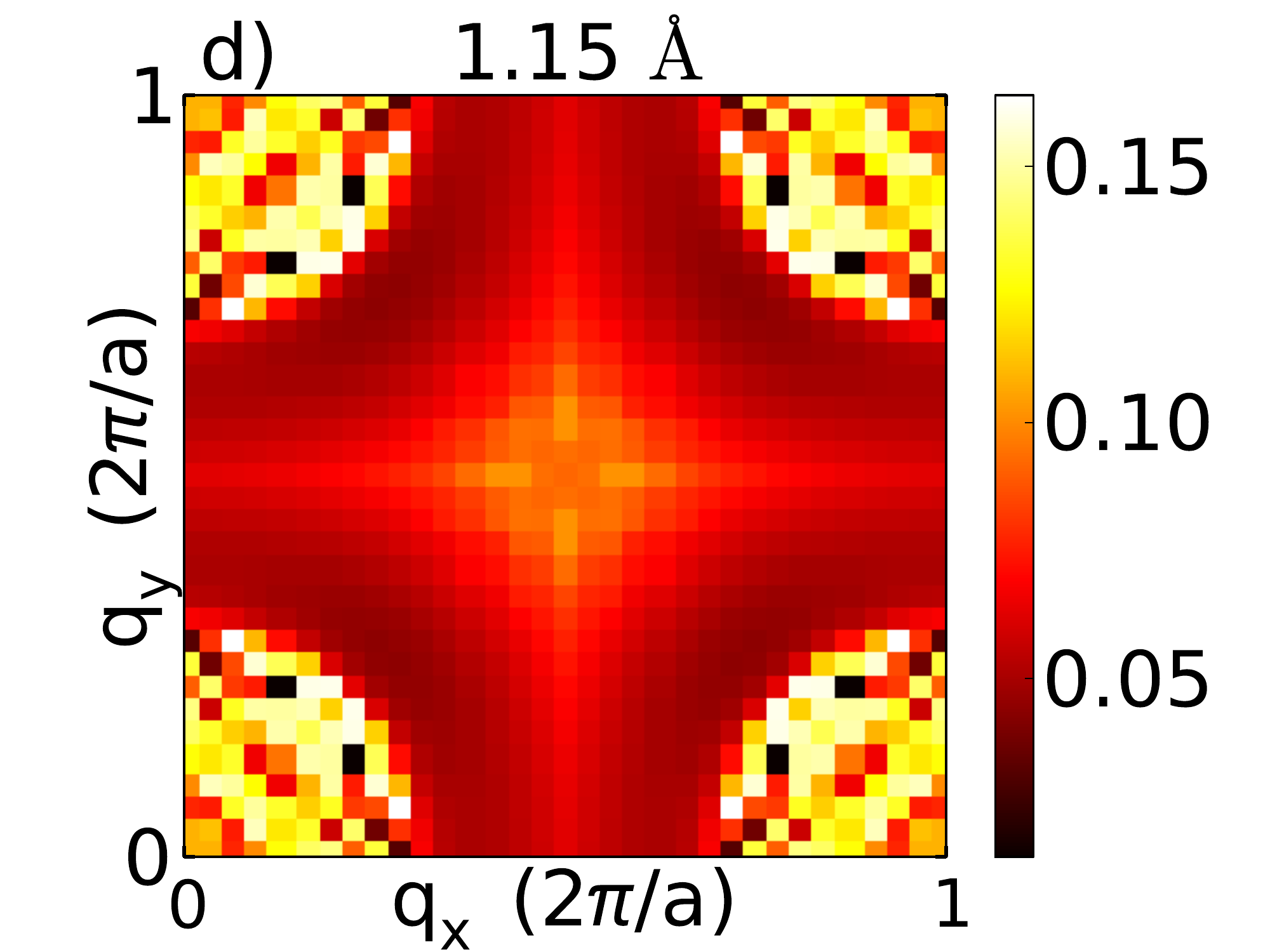} \\
    \includegraphics[width=4.cm]{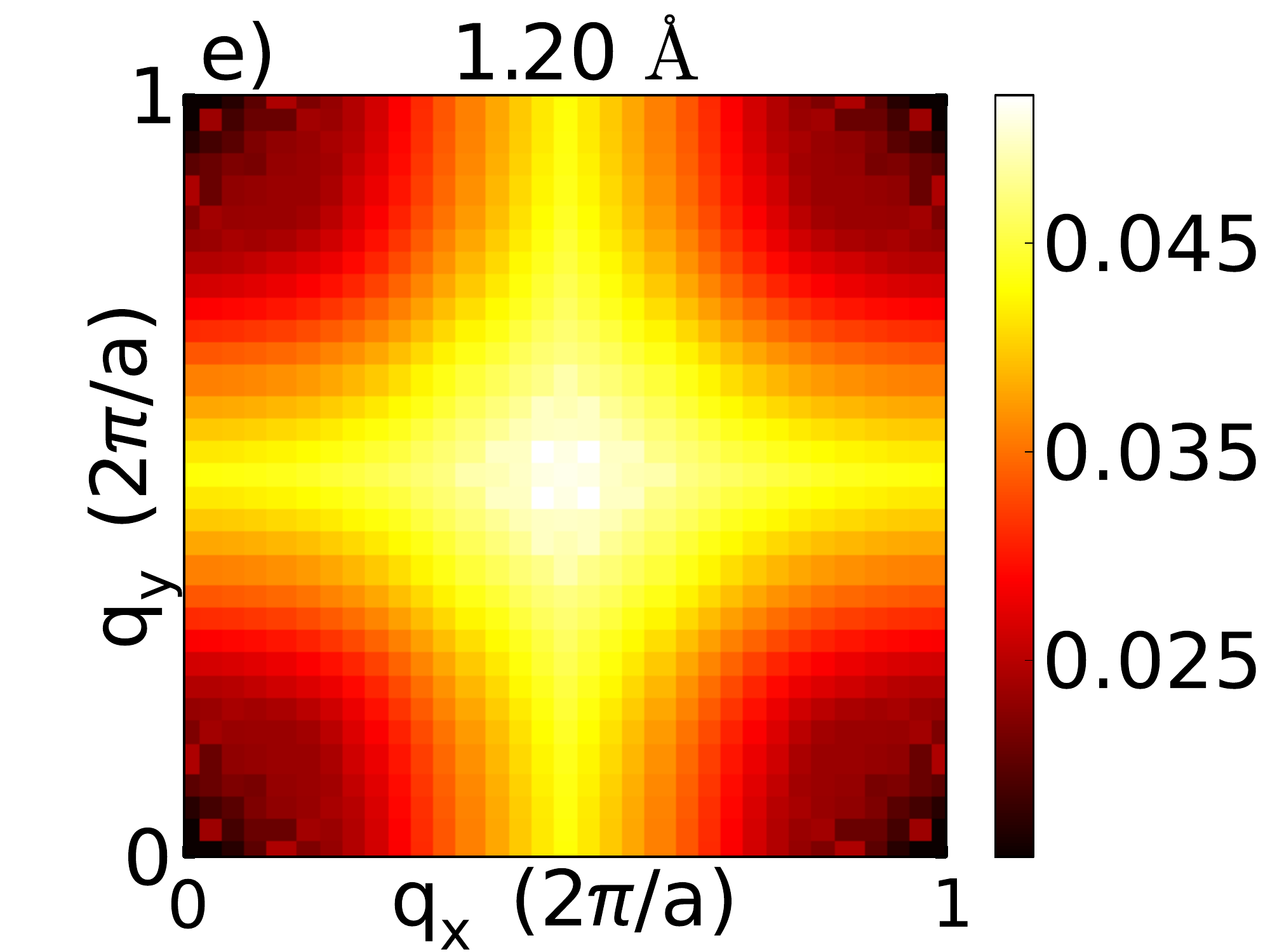} &
  \includegraphics[width=4.cm]{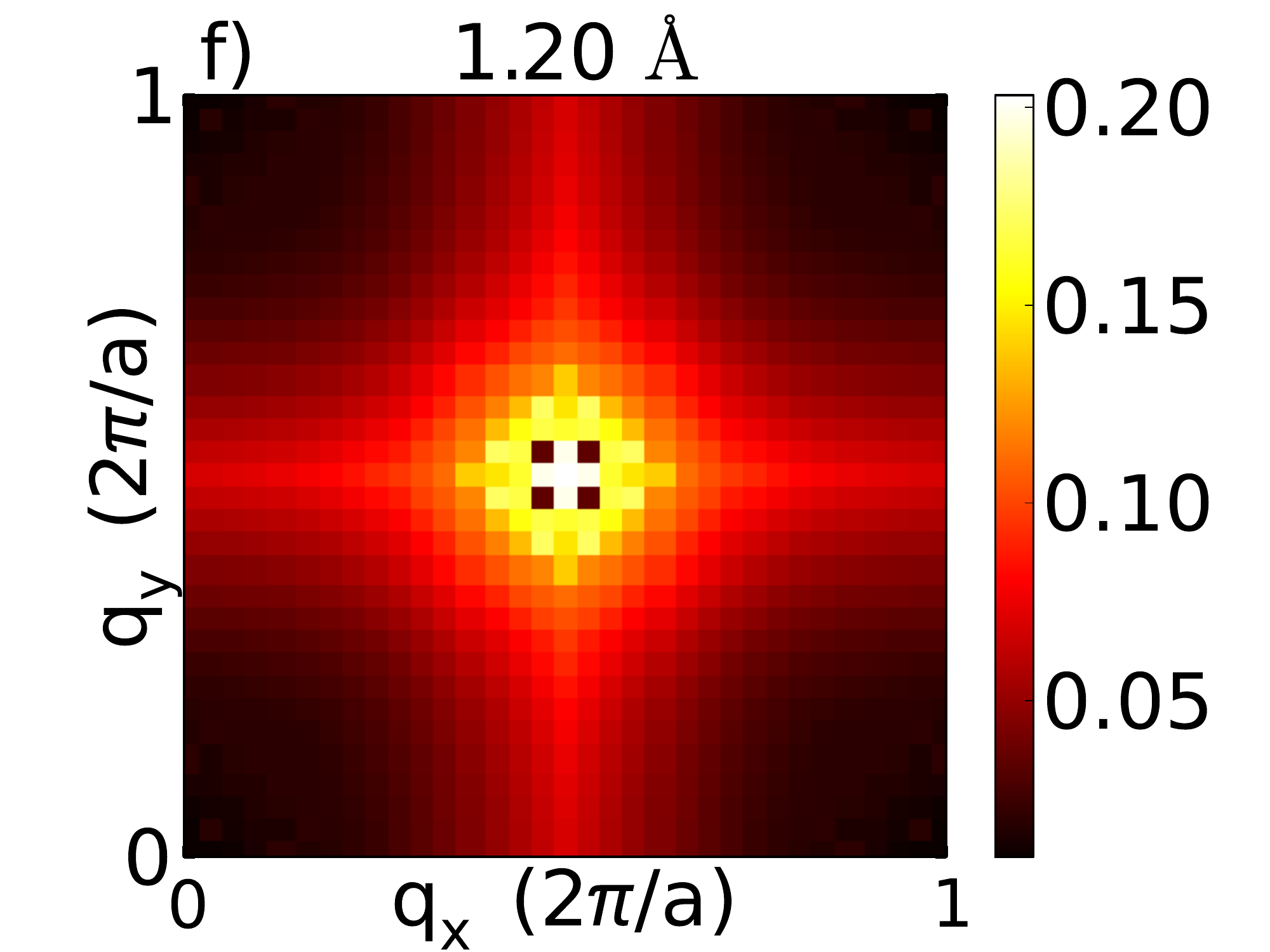} \\
  \end{tabular}
  \caption{Real parts of the static non-interacting [(a), (c), (e)] and
    interacting [(b), (d), (f)] spin susceptibilities of iron selenide for
    selenium heights of 1.10~\AA~[(a) and (b)], 1.15~\AA~[(c) and (d)] and
    1.20~\AA~[(e) and (f)]. Shown is the $\bm{G}=\bm{G}'=0$ component
    of the susceptiblity matrix in Fourier space.}
  \label{fig:chi}
\end{figure}

Figure~\ref{fig:chi} shows the real part of the DFT-LDA static non-interacting
and interacting spin susceptibilities at three Se heights in the
vicinity of the transition from the nonmagnetic phase to the
antiferromagnetic stripe phase. At the smallest Se height (1.10~\AA),
we observe strong features in the non-interacting susceptibility near
$\bm{q}=0$. In the interacting spin susceptibility, these features are
strongly enhanced indicating that at these wave vectors the Stoner
condition, which states that magnetism occurs if $I_{xc}\chi_0(\bm{q})$ (in
matrix notation) has eigenvalues equal to unity, is almost fulfilled.

At a selenium height of 1.15~\AA~[see Figs.~\ref{fig:chi}b) and c)],
another feature at $\bm{q}=(1/2,~1/2,~0)\frac{2\pi}{a}$ emerges in addition to the
structure near $\bm{q}=0$ in the non-interacting susceptibility. This
feature results from spin fluctuations with stripe phase
character. When interactions are included, this new feature, however,
becomes weaker.

Finally, at a selenium height of 1.20~\AA~[see Figs.~\ref{fig:chi}e)
and f)], the system is very close to the transition to the
antiferromagnetic stripe phase, see Fig.~\ref{fig:bands}(a). Now, the
peak at $\bm{q}=(1/2,~1/2,~0)\frac{2\pi}{a}$ becomes very strong and
is further enhanced when interactions are included. Again, the
divergence of the interacting spin susceptibility indicates that the
Stoner criterion is almost fulfilled signaling the onset of a
transition to a stripe phase.

To gain further insight, we fit our results to a model where
off-diagonal elements of the non-interacting susceptibilities in a
plane-wave basis are neglected (neglecting the so-called local-field
effects). We then adjust the Stoner parameter $I_{xc}$, which now is a
single number, until the model reproduces the diagonal elements of the
previously computed interacting spin susceptibility. In this way, we
extract $I_{xc}=0.43$~eV (per Fe atom), which agrees very well with
the Stoner parameter for bcc iron $I_{xc}=0.46$~eV
\cite{janak1977uniform}. In addition, we find that the same value of
$I_{xc}$ can be used for all selenium heights that were studied. This
indicates that the Stoner parameter is only weakly dependent on the
crystalline environment.

\begin{figure}
  \includegraphics[width=8.cm]{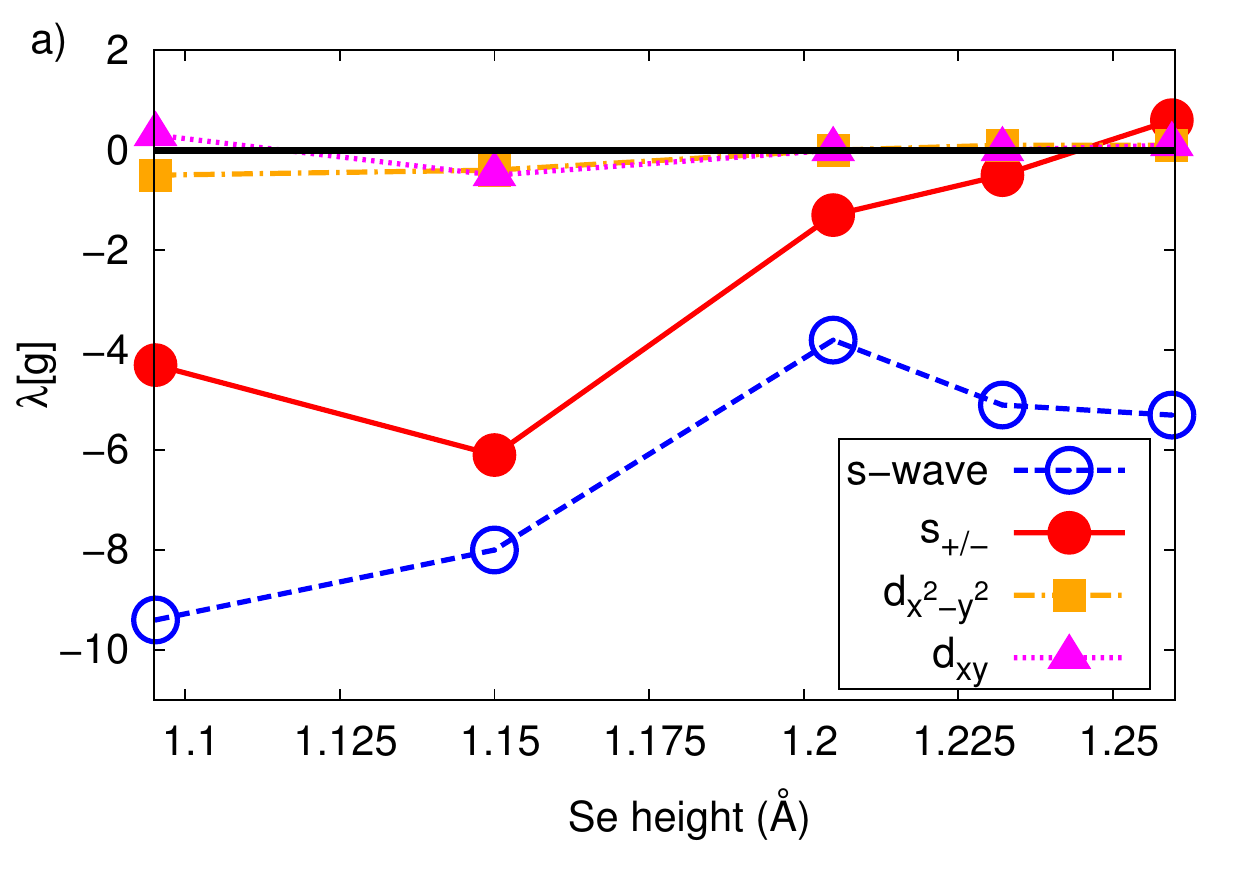}
  \includegraphics[width=6.cm]{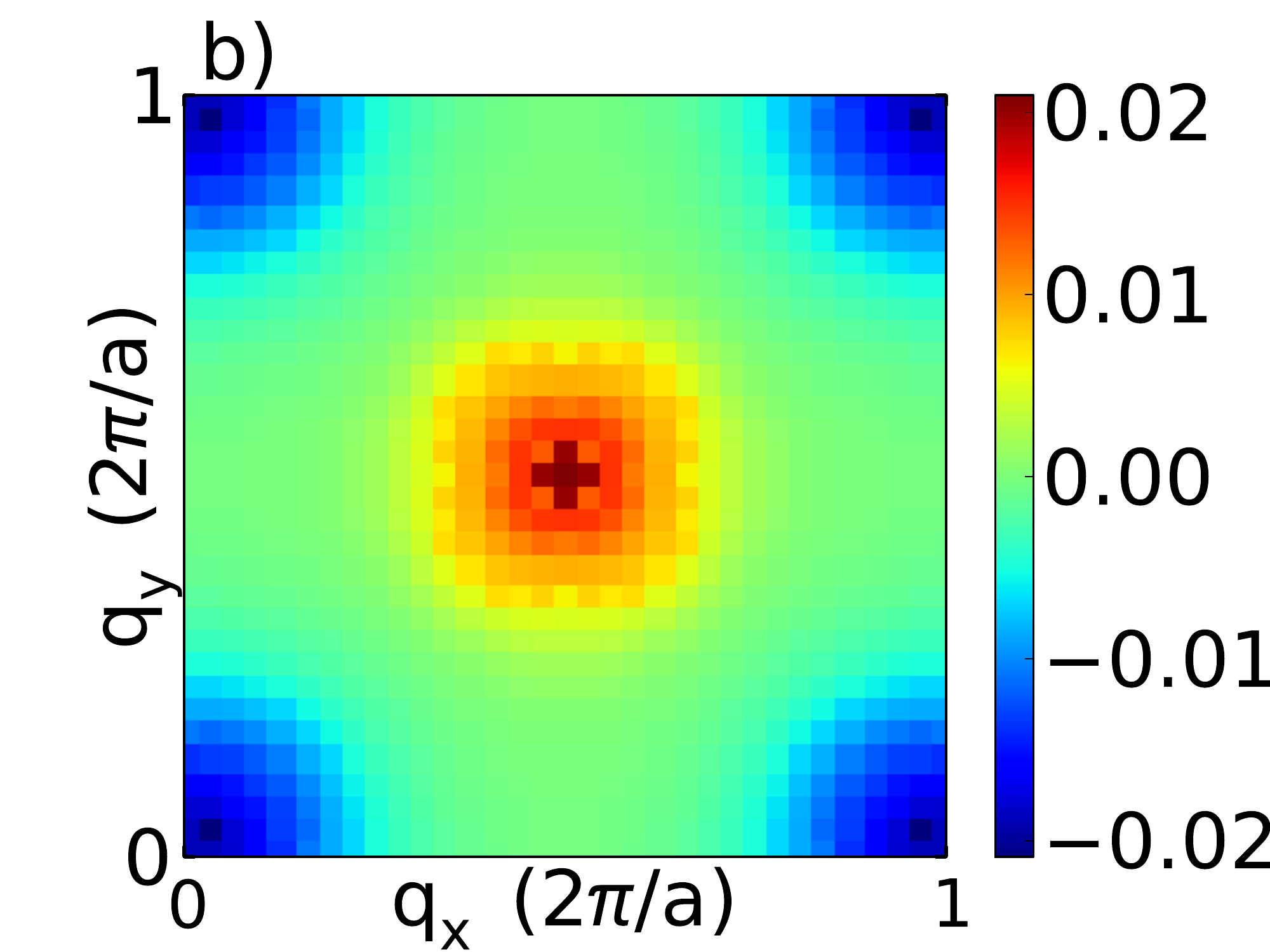}  
  \includegraphics[width=8.cm]{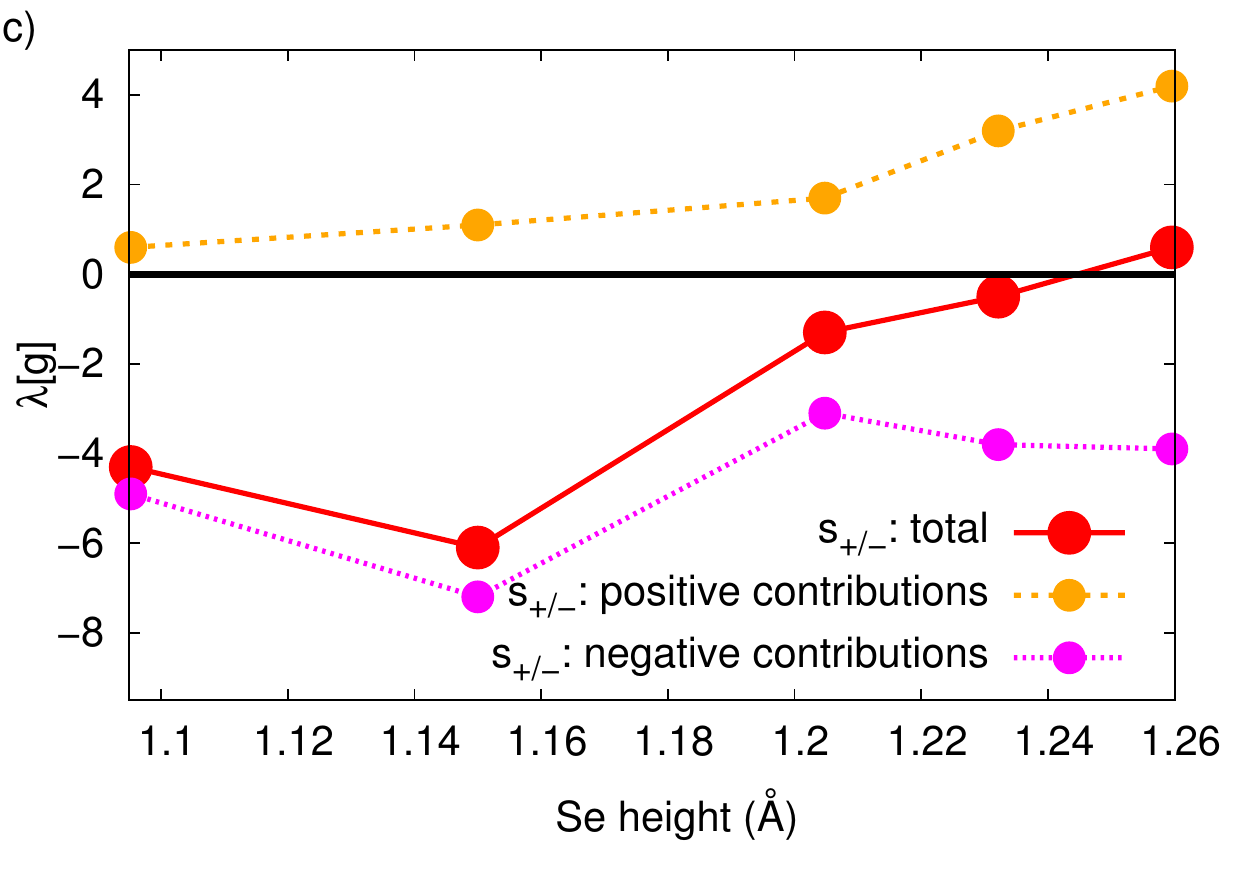}  
  \caption{(a) Electron-spin fluctuation coupling constant for various
  gap symmetries as function of the selenium height. Positive coupling
  constants indicate that the system may undergo a transition to a
  superconducting phase at sufficiently low temperatures. (b)
  Contributions to the total electron-spin fluctuation coupling
  constant for an $s_{\pm}$-symmetric gap from the $\bm{q}$-points in the
  Brillouin zone at a selenium height of 1.20~\AA. (c) Positive and
  negative contributions to $\lambda[s_{\pm}]$ as function of the
  selenium height.}
  \label{fig:lambda}
\end{figure}

Next, we compute the electron-spin fluctuation coupling strength and
evaluate $\lambda[g]$ for the lowest order symmetry functions
$g(\bm{k})$ of the tetragonal crystal. Specifically, we compute the
coupling strength for $s$-wave [$g(\bm{k})=1$], $s_{\pm}$ [$g(\bm{k})
\propto \cos(k_x)+\cos(k_y)$], $d_{x^2-y^2}$ [$g(\bm{k}) \propto
\cos(k_x)-\cos(k_y)$] and $d_{xy}$ [$g(\bm{k}) \propto
\sin(k_x)\sin(k_y)$] symmetries. Fig.~\ref{fig:lambda}(a) shows our
results for the various coupling strengths as function of the selenium
height. In the singlet channel, the spin-fluctuation mediated
interaction between electrons is \emph{repulsive} and therefore the
coupling strength for s-wave symmetry is always negative. The vicinity
to a magnetic phase transition gives rise to an \emph{almost singular}
spin-fluctuation mediated interaction and large, negative coupling
constants $\lambda_s \ll -1$. This repulsion arising from magnetic
interactions is much larger than the repulsion from the screened
Coulomb interactions in standard metals, where $\lambda_s=-\mu \approx
-0.2$ \cite{lee1995first}. Note that this repulsion must also be
overcome by other pairing mechanisms, such as phonons.

Figure~\ref{fig:lambda}(a) shows that d-wave coupling constants are
quite small, but can be positive and thus give rise to
superconductivity at sufficiently low temperatures. Most
interestingly, we observe that the $s_{\pm}$ coupling constant is
large and negative for small selenium heights, but quickly increases
and eventually becomes positive near the transition to a stripe phase
reaching values of approximately unity. We expect that this value of
the $s_{\pm}$ coupling constant is reduced by the other contributions
to the effective interaction (such as the bare Coulomb interaction and
the charge fluctuation-mediated interaction) only by a \emph{small
  amount}, as these contributions are much more isotropic and thus
produce small values when integrated against the anisotropic
$g(\bm{k})$ in Eq.~\eqref{eq:lambda}.

Figure.~\ref{fig:lambda}(b) shows the contributions to the total
$s_{\pm}$ coupling strength from all $\bm{q}$-points in the Brillouin
zone at a selenium height of 1.20~\AA. We observe significant
cancellations between positive and negative
contributions. Fig.~\ref{fig:lambda}(c) shows the total positive and
total negative contributions to the $s_{\pm}$ coupling strength as
function of the selenium height. While the negative contributions
remain relatively constant, the positive contributions increase
rapidly as stripe phase spin fluctuations become enhanced. These spin
fluctuations scatter electrons from the hole pockets at the
$\Gamma$-point to the electron pockets near the M-point. In the
$s_{\pm}$ scenario, the superconducting order paramater switches sign
between the $\Gamma$- and the M-point and therefore such scattering
events are favorable for the emergence of superconductivity. On the
other hand, spin fluctuations with wave vectors in the vicinity of the
$\Gamma$-point scatter Cooper pairs only from electron pockets to
other electron pockets or from hole pockets to other hole
pockets. Their contribution to the $s_{\pm}$ coupling constant is
\emph{negative}. Suppression of such fluctuations (for example, by
application of pressure) could provide a path towards higher
transition temperatures \cite{fernandes2013suppression,imai2009does}.

Finally, we discuss corrections to the presented first-principles
framework that arise from the lack of self-consistency in our
calculations. Specifically, the non-interacting spin susceptibility
should be computed from a quasiparticle band structure including
renormalization effects arising from many-electron interactions
instead of the mean-field DFT-LDA band structure. While such
renormalization effects are reproduced by ``beyond-DFT'' approaches,
such as GW theory \cite{tomczak2012many} or dynamical mean-field
theory \cite{aichhorn2010theoretical}, good agreement with
experimental ARPES measurements can be obtained by simply dividing all
DFT-LDA band energies by a factor $\alpha$,
i.e. $\xi_{n\bm{k}}=\epsilon_{n\bm{k}}-\epsilon_F \rightarrow
\xi'_{n\bm{k}}=\xi_{n\bm{k}}/\alpha$. Based on a recent ARPES
experiment on FeSe multilayers \cite{tan2013interface}, we use
$\alpha=3$, see Fig.~\ref{fig:bands}(b). This rescaling of the band
energies renormalizes the non-interacting susceptibility according to
$\chi'_0=\alpha \chi_0$ and also the density of states at the Fermi
level $D'_F=\alpha D_F$.

In addition to the electronic band structure, also the Stoner
parameter is renormalized in the vicinity of a magnetic phase
transition where spin fluctuations are significant
\cite{aguayo2004n,ortenzi2012accounting}. To include this effect, we
also rescale the Stoner parameter according to
$I'_{xc}=I_{xc}/\beta$. We use $\beta=4.17$ which reproduces the
experimental magnetic moment and critical doping strength in LaOFeAs
\cite{andersen2011multi} (note that $\beta=\alpha \times s$ with
$s=1.39$ being the rescaling parameter for $I_{xc}$ neglecting band
structure renormalization effects).

We now compute the coupling constant for the $s_{\pm}$-symmetric gap
in FeSe at the experimental selenium height employing a renormalized
band structure and Stoner parameter. This yields
$\lambda[s_{\pm}]=0.30$. Including the effect of charge fluctuations
results in a slightly reduced value of $\lambda[s_{\pm}]=0.28$.
A crude estimate of the superconducting
transition temperature $T_c=\omega_{SF} e^{-1/\lambda[s_{\pm}]}$
(where we use $\omega_{SF}=15$~meV, a typical energy scale for spin
fluctuations in iron-based compounds\cite{inosov2009normal}) yields
$T_c = 5$~Kelvin consistent with the experimental transition
temperature of 8 Kelvin in FeSe
\cite{medvedev2009electronic,imai2009does} (note that allowing
$g(\bm{k})$ to be a \emph{sum} of symmetry functions further increases
$\lambda$ \cite{maier2009origin} and our calculation therefore provides
a \emph{lower bound} for the transition temperature). In the current
work, we have not considered the effect of phonons which was recently
shown to be significant in the FeSe monolayer
\cite{coh2014structural}. Future work is needed to investigate the
interplay of spin fluctuations and phonons in iron-based
superconductors.

J. L. acknowledges valuable discussions with Lex Kemper, Sinisa Coh,
Hyungju Oh and Douglas Scalapino. This work was supported by NSF Grant
No. DMR10-1006184 (theoretical analysis) and by the SciDAC Program on
Excited State Phenomena (methods and software developments) and Theory
Program (magnetic susceptibility calculations) funded by the
U. S. Department of Energy, the Office of Basic Energy Sciences and of
Advanced Scientific Computing Research, under Contract
No. DE-AC02-05CH11231 at the Lawrence Berkeley National
Laboratory. Computational resources have been provided by the DOE at
NERSC.

\emph{Note added:} After submission of the present manuscript, a similar
calculation on spin fluctuation-mediated superconductivity in iron
selenide was published \cite{essenberger2014ab}.

\bibliography{paper}
\end{document}